\documentclass[prb,showpacs,twocolumn,floatfix,superscriptaddress]{revtex4}
\usepackage{epsfig}

\begin{document}

\title{g-Factor anisotropy of hole quantum wires induced by the Rashba interaction}
\author{M. Magdalena Gelabert}
\affiliation{Departament de F\'{\i}sica, Universitat de les Illes Balears,
E-07122  Palma de Mallorca, Spain}
\author{Lloren\c{c} Serra}
\affiliation{Departament de F\'{\i}sica, Universitat de les Illes Balears,
E-07122  Palma de Mallorca, Spain}
\affiliation{ Institut de F\'{\i}sica Interdisciplinar i de Sistemes Complexos
IFISC (CSIC-UIB), E-07122 Palma de Mallorca, Spain.}

\date{March 16, 2011}
\revised{July 5, 2011}

\begin{abstract}
We present calculations of the g factors for the lower conductance steps
of 3D hole quantum wires. Our results prove that the 
anisotropy with magnetic field orientation, relative to the wire, originates in the Rashba spin-orbit coupling.
We also analyze the relevance 
of the deformation, as the wire evolves from 3D towards a flat 2D geometry.
For high enough wire deformations, the perpendicular 
g factors are greatly quenched by the Rashba interaction.
On the contrary, parallel g factors are rather insensistive
to the Rashba interaction, resulting in a high g factor anisotropy. 
For low deformations we find a more irregular behavior which hints 
at a sample dependent scenario.

\end{abstract}
\maketitle


\section{Introduction}

Spin-orbit interactions in semiconductor materials offer 
interesting possibilities of spin control in nanostructures.\cite{win03}
Among them, the Rashba interaction that originates in externally applied 
electric fields is most promising due to its tunability. In this work we 
prove that the Rashba interaction is an important source of spin anisotropy
in hole quantum wires. 
This anisotropy manifests in large differences between the energy 
splittings for magnetic fields parallel and perpendicular to the 
wire.\cite{dan97,dan06,kod08,klo09}
Our calculations show that in the presence of Rashba interaction the 
perpendicular field becomes much less effective in generating spin splittings
than the parallel one. This effect is favored
by the deformation of the quantum wire, i.e., anisotropy
increases when the wire evolves from 3D towards a more flat quasi 2D geometry.

In semiconductor hole systems like p-type GaAs nanostructures transport 
is mediated by holes in the valence bands. As compared to electrons,
holes are characterized by a
spin 3/2, besides a sign difference in charge. The corresponding fourfold discrete space is a source of
qualitative differences with respect to the more usual twofold spin of electrons.
In 2D hole gases different splittings for normal and
in-plane fields have been observed, 
as well as for different in-plane orientations.\cite{win00}
By further confining the hole gas it is possible to generate 
nanostructures with the shape of quantum wires. In this case, 
the splitting varies, in principle, with both wire and
magnetic field orientations.\cite{dan06,kod08,klo09} 

There are few theoretical analysis of the 
spin splittings in hole quantum wires.\cite{gol95,har06,cso07} 
Although the Rashba interaction was usually not taken into account,
this situation changed in some recent works.\cite{che11,qua10}
Indeed, Quay {\em et al.}\cite{qua10}
have observed the formation of a spin-orbit
gap induced by the combined action of magnetic field and 
Rashba coupling in a hole quantum wire, while 
Chesi et {\em al.}\cite{che11} have studied, both experimentally and theoretically,
the spin resolved transmission of a quantum point contact 
fabricated in a 2D hole gas. In the latter, the Rashba interaction
is shown to favor a band crossing at finite wavenumber that 
can be manipulated with an external magnetic field. In agreement
with our results, this crossing is obtained in a multiband description 
of the hole states. It can also be explained within a restricted 
single band description adding a cubic Rashba term.

In this work we have focussed our attention on the structure-inversion-asymmetry (Rashba) splitting 
since this is known to be the dominant source of spin-orbit coupling in GaAs.
The bulk-inversion-asymmetry (Dresselhaus) is much smaller and has a minimal
effect on the energy bands.\cite{qua10,win00}
We will show  that in a hole quantum wire
oriented along $x'$ 
the Rashba interaction due to asymmetry in 
the growth direction ($z'$) causes a large difference
between parallel ($x'$) and perpendicular ($y'$) g factors
of the wire, as deduced from the $B$-induced splittings of the 
conductance steps. 
This anisotropy is due to the quenching of the 
splitting when $B$ is along $y'$ and the wire flatness is large.
For smaller deformations the situation is less clear due to
a non monotonous evolution of the splittings that may result 
in a sample-dependent scenario. 

\section{Model}

We describe the anisotropic kinetic energies  
${\cal H}^{\it (kin)}$ of the holes
in a 4-band kp 
model. Introducing a spin discrete index 
$\eta=3/2,\dots,-3/2$ and following the notation of
Ref.\ [\onlinecite{win03}] the diagonal terms read
\begin{equation}
{\cal H}^{\it (kin)}_{\eta\eta}
=
-\frac{\hbar^2}{2m_0}
\left[
(\gamma_1+c_\eta \gamma_2) k_\parallel^2
+
(\gamma_1-2c_\eta\gamma_2)k_z^2
\right]\; ,
\label{eq1}
\end{equation}
where $c_{\pm 3/2}=1$ and $c_{\pm 1/2}=-1$. In Eq.\ (\ref{eq1})
$\gamma_1$ and $\gamma_2$ are the kp parameters, $\vec{k}$ is the 
3D wavenumber and we have also defined
$k_\parallel^2=k_x^2+k_y^2$. 
The nondiagonal kinetic terms are
\begin{eqnarray}
{\cal H}^{\it (kin)}_{+\frac{3}{2},+\frac{1}{2}}&=&\frac{\hbar^2}{m_0}\sqrt{3}\,\gamma_3\,k_-k_z\; ,\nonumber\\
{\cal H}^{\it (kin)}_{+\frac{3}{2},-\frac{1}{2}}&=&\frac{\hbar^2}{2m_0}\sqrt{3}(\gamma_2\hat{K}-2i\gamma_3 k_x k_y)\; ,\nonumber\\
{\cal H}^{\it (kin)}_{+\frac{1}{2},-\frac{3}{2}}&=&{\cal H}^{\it (kin)}_{+\frac{3}{2},-\frac{1}{2}}\; ,\nonumber\\
{\cal H}^{\it (kin)}_{-\frac{1}{2},-\frac{3}{2}}&=&-{\cal H}^{\it (kin)}_{+\frac{3}{2},+\frac{1}{2}}\; ,
\end{eqnarray}
where 
$k_\pm=k_x\pm ik_y$
and 
$\hat{K}=k_x^2-k_y^2$.
We only refer to contributions in the upper triangle of 
matrix ${\cal H}^{\it (kin)}_{\eta\eta'}$
since the remaining ones
can be inferred from the Hermitian character of the matrix.
In all calculations discussed below we have used numerical values for the 
kp parameters $\gamma$'s corresponding to GaAs.\cite{win03}

The wire confinement is represented by a deformed 2D harmonic oscillator.
Assuming the wire is oriented along $x'$ while transverse and
growth directions are given by $y'$ and $z'$, respectively, it is
\begin{equation}
\label{eq3}
{\cal H}^{\it (conf)}
= -\frac{1}{2}m_0\omega_0^2({y'}^2 + a {z'}^2)\; .
\end{equation}
The adimensional parameter $a$ of Eq.\ (\ref{eq3}), corresponding to
the ratio of confinement strengths in $z'$ and $y'$, 
controls the flatness or 2D character of the wire.  
The direct coupling with the magnetic field $\vec{B}$ is given by the Zeeman
term
\begin{equation}
\label{eqZ}
{\cal H}^{(Z)}
= -2\kappa\mu_B \vec{B}\cdot\vec{J}
\; ,
\end{equation}
where $\kappa$ is a kp parameter, $\mu_B$ represents the Bohr magneton and $\vec{J}$ 
is the angular momentum operator for a spin 3/2.
Finally,
the Rashba interaction is described by 
\begin{equation}
{\cal H}^{(R)}
= (\vec{k}\times\vec{\cal R})\cdot\vec{J}
\; ,
\end{equation}
where we defined a vector constant $\vec{\cal R}\equiv\alpha\vec{\cal E}$, related to the 
effective electric field $\vec{\cal E}$ and kp parameter $\alpha$.\cite{win03}
We shall treat $\vec{\cal R}$ as a two-parameter vector with dominant component 
along the growth direction, i.e.,
$\vec{\cal R}={\cal R}_{z'}\hat{u}_{z'}+{\cal R}_{y'}\hat{u}_{y'}$ with 
${\cal R}_{z'}>{\cal R}_{y'}$.

In the presence of a magnetic field, the orbital effects of the field are taken into account 
by means of the substitution
$\vec{k} \to -i\nabla-\frac{e}{\hbar c}\vec{A}$ with the 
vector potential $\vec{A}=(-yB_z+zB_y,-B_xz/2, B_xy/2)$. In this process, Hermiticity
is enforced in the cross terms by using the symmetrized forms, such as
$k_xk_y\to(k_xk_y+k_yk_x)/2$.
Summarizing all contributions the total Hamiltonian reads
\begin{equation}
\label{hfull}
{\cal H}\equiv {\cal H}^{\it (kin)}+{\cal H}^{\it (conf)}+{\cal H}^{(Z)}+{\cal H}^{(R)}\; .
\end{equation}
The wire Hamiltonian eigenvalues can be labelled with $q$, a real number 
representing the longitudinal momentum and an index $I=1,2,\dots$ as
\begin{equation}
\label{eq7}
{\cal H}(q)|Iq\rangle=\varepsilon_I(q)|Iq\rangle\; ,
\end{equation}
where $\varepsilon_I(q)$ are the discrete energy bands of the nanostructure.
The eigenvalues are ordered as 
$\varepsilon_1(q)\ge\varepsilon_2(q)\ge\dots$ since the spectrum is not
bounded from below due to the negative kinetic terms.

We have obtained the solutions of the eigenvalue problem 
given by Eq.\ (\ref{eq7})  
by discretizing in harmonic oscillator states for the two transverse
oscillators along $y'$ and $z'$,
\begin{equation}
\label{eq8}
|Iq\rangle = \sum_{nm\eta}{C_{nm\eta}^{(Iq)}}\,|nm\eta\rangle\; ,
\end{equation}
where $n,m=0,1,\dots$ represent the number of quanta in each oscillator,
respectively.
The resulting matrix eigenvalue problem reads
\begin{equation}
\sum_{nm\eta}
{\langle n'm'\eta'|{\cal H}(q)|nm\eta\rangle\,
C^{(Iq)}_{nm\eta}
=
\varepsilon_{I}(q)\,
C^{(Iq)}_{n'm'\eta'}}
\; .
\end{equation}

In practice the number of oscillator states 
in expansion Eq.\ (\ref{eq8})
can be truncated once convergence of the results
is ensured. The results shown below are well converged and they have been obtained
including the lower 20 oscillator 
states in each direction. In Appendix \ref{appB} a precise discussion on the 
relevance of the basis truncation is given.

\section{Results and discussion}

\begin{widetext}

\begin{figure}[h]
\epsfig{file=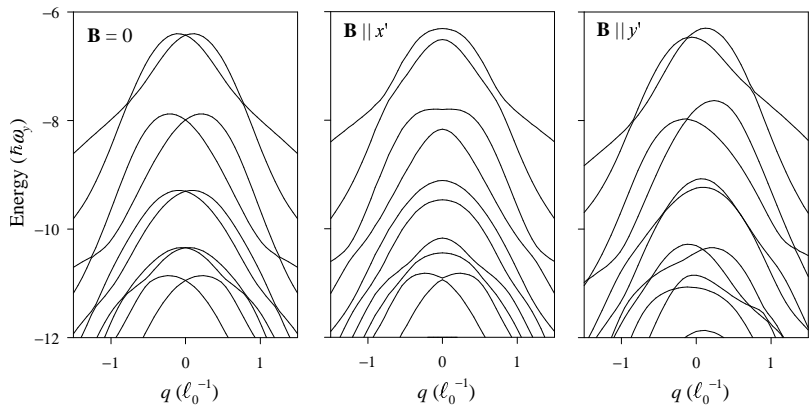,angle=0,width=0.6\textwidth,clip}
\caption{Energy bands for $a=64$,
${\cal R}_{z'}=2.6\hbar\omega_0\ell_0$ and ${\cal R}_{y'}=0$. 
Left panel is for $B=0$ while
center and right ones are for $\mu_B B=0.1\hbar\omega_0$ in the parallel and
transverse directions, respectively. The wire is oriented
along $(-2,3,3)$ and the growth direction is $(3,1,1)$.
}
\label{fig1}
\end{figure}

\end{widetext}

As illustrative examples, Fig.\ \ref{fig1} displays the energy bands
of selected cases.
As is well known, the Rashba interaction causes a 
characteristic 
band structure easily recognizable by the pairs of subbands crossing at
$q=0$ and with maxima at opposite $q$ values (left panel). 
These maxima correspond to band energy minima for the case of electrons.   
In the presence of a magnetic field, when this points along the 
wire ($x'$, central panel), an anticrossing of the bands appears at $q=0$. 
This anticrossing may lead to anomalous conductance steps,
similar to those recently measured in
Ref.\ \onlinecite{qua10}. In Fig.\ \ref{fig1} this behavior can be 
seen for $(E,q)\approx (-11\hbar\omega_0,0)$.
For $B$ in the transverse direction ($y'$, right panel)
the band crossings persist, but the two central maxima 
for each pair of bands are shifted differently 
in energy, the band structure becoming asymmetric with respect to $q$ inversion.

The $B$-induced modifications of the band structure, as seen in Fig.\ \ref{fig1},
cause a change in the conductance of the wire. This modification of the
conductance, in the limit of weak magnetic field, is conveniently summarized
by a number called the g factor of each conductance split level. At $B=0$, time reversal invariance
of the system causes the conductance $G$ to increase in steps 
of $2G_0$ as the Fermi 
energy of the leads is reduced, where $G_0=e^2/h$ is the conductance quantum.
The evolution of the wire conductance with energy
can be understood 
if we imagine a horizontal line, indicating
the position of the Fermi energy, in the left panel of Fig.\ 1; as this line
is moved to lower energies it sweeps the band maxima always in pairs, each maxima 
corresponding to an increase of $G_0$ in the conductance for hole transport.
The result is the typical staircase conductance, with step heights
of $2G_0$.
A similar procedure for the central and right
panels of Fig.\ \ref{fig1} convince ourselves that  
intermediate half steps in conductance are caused by the magnetic field.
They are smaller than the full steps and proportional to the intensity 
of the magnetic field.

\begin{figure}[t]
\epsfig{file=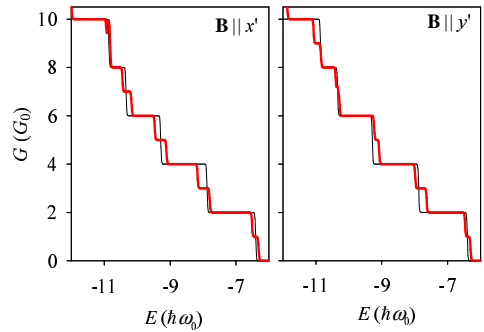,angle=0,width=0.35\textwidth,clip}
\caption{(Color online) 
Conductance traces for the band structure in parallel and perpendicular magnetic fields 
of Fig.\ \ref{fig1}. For comparison, the $B=0$ conductance is
displayed as a thin line.
}
\label{condfig}
\end{figure}

The scenario we have just sketched is explicitly shown in Fig.\ \ref{condfig},
highlighting the conductance half steps at odd multiples of $G_0$.
Notice that the energy span varies for each specific conductance half step. 
In the limit of weak magnetic fields we can 
conveniently summarize the $B$-induced $N$th half step in the conductance,
appearing between steps at $2(N-1)G_0$ and $2NG_0$, in 
terms of a single number called the g factor. As this number depends
on the conductance step and the magnetic field orientation, we use
the notation $g^{(N)}_\parallel$ and $g^{(N)}_\perp$ to indicate the 
g factor of the $N$th step, for $B$ along $x'$ and $y'$, respectively.
Of course, other orientations are in principle possible, but we will 
restrict first to these two as they are the relevant ones in the measurements
of spin hole anisotropy. In Appendix \ref{appA} we will briefly mention 
the behavior for $z'$-oriented field.

Our precise definition of the parallel-field g-factor is 
\begin{equation}
\label{eq10}
g^{(N)}_{\parallel} =
\frac{\Delta^{(N)}_\parallel}{3 \mu_B B}\; 
\end{equation}
where $\Delta_\parallel^{(N)}$ is the energy range for the $N$th half
step in a magnetic field $B$. In Eq.\ (\ref{eq10}), the factor 3 in the 
denominator  is introduced by
convention.\cite{gg} The definition of 
$g^{(N)}_\perp$, for magnetic field along $y'$, is obtained simply replacing 
$\Delta_\parallel^{(N)}$ by $\Delta_\perp^{(N)}$
in Eq.\ (\ref{eq10}).

\begin{widetext}

\begin{figure}[t]
\epsfig{file=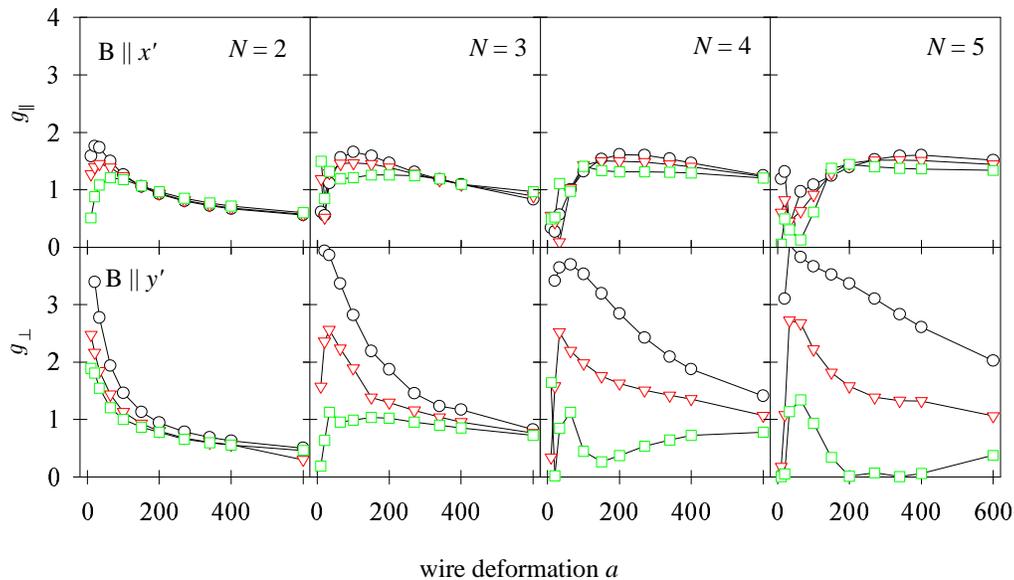,angle=0,width=0.75\textwidth,clip}
\caption{(Color online) Parallel and perpendicular g factors
as a function of wire deformation for different values of the Rashba 
strength:  ${\cal R}_{z'}=0$ (circles), $1.5\hbar\omega_0\ell_0$ (triangles)
and $2.6\hbar\omega_0\ell_0$ (squares).
Upper and lower rows are for parallel and perpendicular fields while 
columns from left to right correspond to increasing conductance half step $N$ (see text).
The results for $N=1$ are not shown due to their similarity with the displayed 
$N=2$ case. The orientation of the wire is the same of Fig.\ \ref{fig1}. 
}
\label{fig2}
\end{figure}

\end{widetext}

\begin{figure}[b]
\epsfig{file=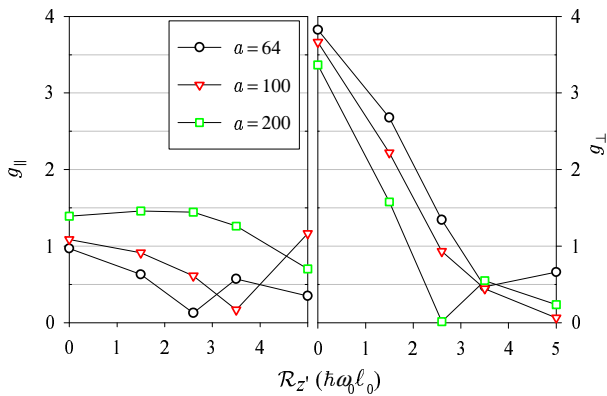,angle=0,width=0.45\textwidth,clip}
\caption{(Color online) 
Parallel (left) and perpendicular (right) g factors as a function of 
Rashba coupling intensity for different values of the deformation $a$.
The results correspond to the $N=5$ conductance half step.
}
\label{fig3}
\end{figure}

Figure \ref{fig2} displays the perpendicular (lower row) and parallel (upper row) g factors
for the lower conductance steps, as a function of the wire deformation $a$
and for different values of the Rashba coupling ${\cal R}_{z'}$. 
These are the main results of our work.
They were obtained for a specific wire orientation and
direction of crystallographic growth ($z'$) taken 
from the experimental works of Danneau {\em et al.}\cite{dan06} 
and Koduvayur {\em et al}.\cite{kod08}
We have checked, however, that a
qualitatively similar influence of the Rashba intensity and confinement deformation 
are obtained assuming other arbitrary orientations. 
The g factors show a general tendency to decrease as $a$ increases,
except for smaller deformations ($a<100$) for which $g$ may increase 
or even show irregular behavior in some cases.
Focussing first on $g_\parallel$, we notice that this component 
does not change significantly when the
Rashba intensity increases, specially at large $a$'s, for which the 
results are almost overlapping
in the upper panels of Fig.\ \ref{fig2}.
Very remarkably, however, 
for magnetic field in the perpendicular direction
small variations in ${\cal R}_{z'}$ are enough to strongly 
modify the values of $g_\perp$. This is more clearly seen in Figure \ref{fig3},
which displays the dependence with Rashba coupling intensity of the g factors. 

There is a general Rashba-induced quenching
of $g_\perp^{(N)}$ in Figs.\ \ref{fig2} and \ref{fig3}, quite conspicuous for $N=4$ and 5. This effect
is so strong that it can reverse the relative importance of $g_\parallel$ and
$g_\perp$; from $g_\perp > g_\parallel$ when ${\cal R}_{z'}=0$ to
$g_\perp  << g_\parallel$ for increasing ${\cal R}_{z'}$ ($> 2.5\hbar\omega_0\ell_0$, Fig.\ \ref{fig3}).
With the chosen values of ${\cal R}_{z'}$ we even find a range 
of $a$'s for which $g_{\perp}^{(5)}$ essentially vanishes.
It is interesting to point out that 
a similar quenching of conductance plateaus in transverse field
was discussed in Ref.\ \onlinecite{ser05} for parabolic wires
with electron conduction, as opposed to the present hole conduction. 
In both cases the Rashba spin-orbit coupling is the underlying mechanism.  

Turning to the comparison with experiments, this is somewhat complicate
due to the sample dependence. In general, however, 
a large g-factor anisotropy between parallel and perpendicular 
orientations has indeed been observed in Refs.\ \onlinecite{dan06,kod08,klo09}. 
This was generally attributed to a preferential orientation of the spins along the 
wire for strong confinements. Our results prove with detailed calculations 
that the Rashba interaction for holes
is the specific mechanism allowing the appearance of this anisotropy.  As this
interaction is sample dependent and may vary with external field, our results 
also predict that the hole g factors may be tunable to a certain degree,
what may be relevant for spintronic applications. 
The experimental values 
of wire deformation $a$ are somewhat uncertain in general, which is 
an additional source of difficulty for comparison. 
In general, however, experimental wire deformations 
are $a<100$, which in our calculations corresponds to a regime
with rather large fluctuations (Fig.\ \ref{fig2}). Only for larger $a$'s
the value of
$g_\perp^{(N)}$ is consistently below $g_\parallel^{(N)}$ at
high enough ${\cal R}_{z'}$. We believe that detailed comparison in this 
regime is quite involved due to the fluctuations. On the other hand, these
sharp variations of $g_\parallel$ in the small-$a$ regime and of
$g_\perp$ at all $a$'s can be seen as a manifestation of 
magnetoconductance tunability via the Rashba coupling.  

\section{A two-band model}

A more transparent physical interpretation, complementing the above numerical 
results, can be obtained in a simplified model based on only two bands. 
Focussing on the $I$-th intermediate half step having conductance $IG_0$, 
with $I=1,3,\dots$, we select the two states $I$ and $I+1$ at a given 
$q$, $\{ |Iq\rangle_0,|(I+1)q\rangle_0\}$, where the zero subscript is indicating
absence of a magnetic field. These two states are the basis in which the 
effect of the magnetic field in different orientations will be described.

Let us assume that the $B$-field Hamiltonian may be split as
\begin{equation}
\label{h2b}
{\cal H}(q)={\cal H}_0(q)+{\cal H}^{(Z)}\; ,
\end{equation}
where ${\cal H}^{(Z)}$ is the Zeeman energy defined above in Eq.\ 
(\ref{eqZ}) and ${\cal H}_0$ is the zero field Hamiltonian in Eq.\ (\ref{hfull}). 
Notice that 
Eq.\ (\ref{h2b}) neglects orbital field effects, a simplifying assumption
motivated by the qualitative nature of the present two-band model.

The zero-field energy bands, given by
\begin{equation}
{\cal H}_0(q) |Iq\rangle_0 = \varepsilon_{I0}(q)|Iq\rangle_0\;,
\end{equation}
are assumed known; such as those displayed in the left panel of Fig.\ \ref{fig1}
for a specific confinement and Rashba intensity. In presence of a 
magnetic field the modified energy bands are the eigenvalues of the 
matrix
\begin{equation}
\label{m2b}
\left(
\begin{array}{cc}
\varepsilon_{I0}(q) + \gamma_I & \delta \\
\delta^* & \varepsilon_{(I+1)0}(q) + \gamma_{I+1}
\end{array}
\right)\; ,
\end{equation}
where
\begin{eqnarray}
\gamma_I &=& _0\!\langle Iq\vert{\cal H}^{(Z)}\vert Iq\rangle_0 \nonumber\\
\delta &=& _0\!\langle Iq\vert{\cal H}^{(Z)}\vert (I+1) q\rangle_0\; .
\label{gd2b}
\end{eqnarray}

\subsection{Parallel field}
\begin{figure}[t]
\epsfig{file=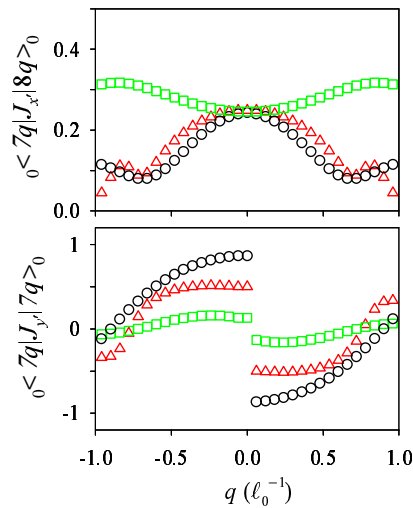,angle=0,width=0.3\textwidth,clip}
\caption{(Color online) 
Wavenumber dependence of the matrix elements entering Eq.\ (\ref{gpa2b})
and (\ref{gp2b}) for $I=7$.
Circles, triangles and squares are for increasing values of the Rashba intensity
${\cal R}_{z'}=0.1\hbar\omega_0\ell_0$, 
$1.5\hbar\omega_0\ell_0$ and 
$2.6\hbar\omega_0\ell_0$, 
respectively. (Other parameters: $a=150$, $B=0$.)}
\label{fig4}
\end{figure}

In a parallel field ${\cal H}^{(Z)}\propto J_{x'}$ and, for this case, we have 
found that the $\gamma_I's$ vanish. This is reminiscent of the 
behavior of conduction 
electron wires, where the spin textures also show a vanishing integrated 
spin along the wire.\cite{ser05} In parallel orientation the band extrema
are at $q=0$ (see Fig.\ \ref{fig1}, middle panel) for which
$\varepsilon_{I0}(0)=\varepsilon_{(I+1)0}(0)$
due to Kramers degeneracy. Under these conditions we find
from the two eigenvalues of the matrix in Eq.\ (\ref{m2b}) that
\begin{equation}
\label{gpa2b}
g_\parallel^{(I)}
=
\frac{4}{3}\kappa 
\left|\rule{0cm}{0.4cm}\,
_0\!\langle I0|J_{x'}|(I+1)0\rangle_0
\right|\; .
\end{equation}
That is, the parallel g-factor is determined by the transition matrix elements
of the parallel spin component between the Kramers degenerate states
at $q=0$. The upper panel of Fig.\ \ref{fig4} shows this transition matrix element
for $I=7$. Notice that for $q\approx 0$ the transition matrix element is not
depending on the Rashba intensity, thus explaining why the parallel g-factor 
is not strongly affected by the spin-orbit coupling. 

\subsection{Perpendicular field}

For ${\cal H}^{(Z)}\propto J_{y'}$ the band maxima are shifted in opposite directions
for positive and negative $q$'s (Fig.\ \ref{fig1} right panel). This implies that the 
energy difference determining the g factor corresponds now to states with opposite 
wavenumbers, say $q_m$ and $-q_m$. For nonzero $q_m$ the two states 
$\varepsilon_{I0}(q_m)$
and 
$\varepsilon_{(I+1)0}(q_m)$
are nondegenerate and, for a sufficiently small field, we should have 
$\delta<<\varepsilon_I(q_m),\gamma_I$ 
in Eq.\ (\ref{m2b}). As a matter of fact, we find that $\delta$ actually vanishes
for the perpendicular field. 
This is the regime of non-degenerate first-order
perturbation theory with modified energies
$\varepsilon_{I0}(q_m)+\gamma_I$ and  $\varepsilon_{(I+1)0}(q_m)+\gamma_{I+1}$.
With the explicit definition of the $\gamma$'s 
and noting that 
$\varepsilon_{I0}(q_m)=\varepsilon_{I0}(-q_m)$
and
$\gamma_I(q)=-\gamma_I(-q)$ for any $q$ (Fig.\ \ref{fig4})
the perpendicular
g factor reads
\begin{equation}
\label{gp2b}
g_\perp^{(I)}
=
\frac{4}{3}\kappa 
\left|\rule{0cm}{0.4cm}\,
_0\!\langle Iq_m|J_{y'}|Iq_m\rangle_0
\right|\; .
\end{equation}

It seems natural that in $y'$ orientation the g factor is simply proportional 
to the expectation value of $J_{y'}$. Figure \ref{fig4} shows the variation of this 
expectation value with the wavenumber and the Rashba intensity. Notice that  
typically $-0.5< q_m < 0.5$, i.e., the maxima are located in the 
central part of Fig.\ \ref{fig4} lower panel. 
When the Rashba intensity ${\cal R}_{z'}$ increases 
there is a severe reduction of $\langle J_{y'}\rangle_0$ in absolute value for this central 
region. This is the mechanism by which the Rashba interaction quenches
the transverse g factor; namely, by means of a strong reduction of the transverse 
$y'$ spin component. 

For strong spin-orbit coupling the expectation values of all three components 
of the spin vector at zero magnetic field, $\langle\vec{J}\rangle_0$, vanish; 
a manifestation of the spin randomization induced by the Rashba field
$\vec{\cal R}$. In $y'$ orientation this induces a quenching of the 
g factor through Eq.\ (\ref{gp2b}) but, quite remarkably, Kramers
degeneracy at zero wavenumber keeps the parallel g factor almost 
unaffected by virtue of the transition matrix elements in Eq.\ (\ref{gpa2b}).

The g factors obtained from Eqs.\ (\ref{gpa2b}) and (\ref{gp2b})
nicely agree with the results from the full diagonalization when orbital
effects of the magnetic field are also neglected in the latter. The comparison
with the complete model, results of Fig.\ \ref{fig2}, is less 
good; the trends are qualitatively reproduced but differences may be as
large as a factor two. Orbital effects of the field are thus quite important for 
a precise analysis.

\section{Conclusion}

We have attributed the anisotropy of 
magnetotransport g factors
in hole quantum wires 
to the Rashba interaction. When the wire deformation and Rashba interaction 
are both large enough ($a>100$, ${\cal R}_{z'}>2.5\hbar\omega_0\ell_0$)  
$g_\perp^{(N)}$ is greatly quenched by the Rashba interaction 
and $g_\parallel^{(N)}$ is almost unaffected. For lower wire deformations
($a<100$) we find a fluctuating, sample dependent behavior of the 
g factors.

\acknowledgments
This work was supported by grant No.\ FIS2008-00781 from MICINN.

\appendix

\section{Field along $z'$}
\label{appA}

Experimental g factors are usually obtained for magnetic fields
in the $x'y'$ plane, either in parallel ($x'$) or perpendicular ($y'$) direction 
with respect to the wire.
For completeness, in this Appendix we discuss in a qualitative way the effects of the magnetic field 
when this points along the growth direction $z'$.
The energy bands are similar to those of the  $x'$ orientation (middle panel of Fig.\ \ref{fig1}):
they are symmetric respect to $q$-inversion, with anticrossing points at $q=0$,
although the $B$-induced splitting is much stronger.
This enhancement agrees with experiments\cite{dan97} and is surely due to 
the important orbital motions induced by the field in this geometry.
We thus obtain $g_{z'}>g_\parallel$, where $g_{z'}$ and $g_\parallel$ 
denote the g factors for $z'$ and $x'$ fields, respectively. 

Looking at the Rashba-field dependence, $g_{z'}$ behaves similarly 
to $g_{\perp}$ (along $y'$): it decreases with 
increasing $\mathcal{R}_{z'}$ but does not vanish for the maximum value we have taken 
($2.6\hbar\omega_0\ell_0$). 
For strong wire deformation the saturation value corresponds to $g_{z'}\approx5$,
while for in-plane magnetic field it corresponds to $g_{\parallel,\perp}\approx 1.5$ (see Fig.\ \ref{fig2}).
For small values of $a$ the behavior of $g_{z'}$ is less regular, as for the 
other orientations, but it tends to increase
with $a$.
Within the two-band model of Sec.\ IV
we expect
\begin{equation}
g_{z'}^{(I)}=\frac{4}{3}\kappa \bigg| \,_0\langle I0|J_{z'}|(I+1)0\rangle_0\bigg|\;,
\end{equation}
which is equivalent to Eq.\ (\ref{gpa2b}), replacing $J_{x'}\rightarrow J_{z'}$, 
and is now depending on the value of the Rashba intensity.

\section{basis truncation}
\label{appB}

This Appendix discusses the relevance of the truncation of the 
number of oscillator states for the $y'$ and $z'$ oscillators. It is usually assumed that 
the confinement allows the truncation to the lowest, or few lowest, states. Here we 
explicitly check this quantitatively for selected values of $a$, the ratio of the two confinement
strengths.
We restrict, for simplicity, to the $B=0$ case with strong Rashba coupling in the growth
direction. 

\begin{widetext}

\begin{figure}[t]
\epsfig{file=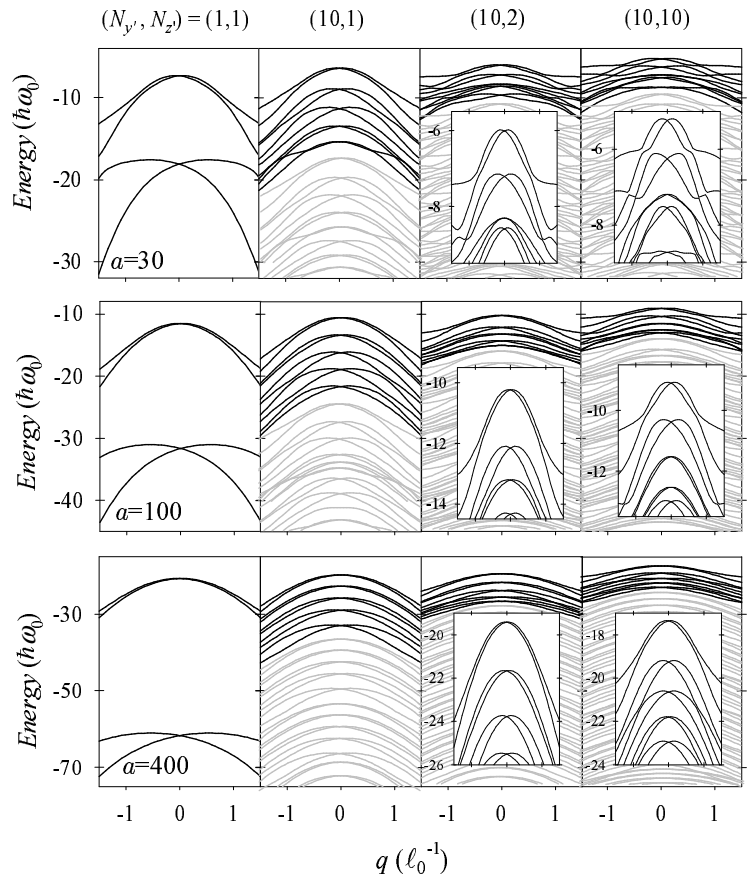,angle=0,width=0.55\textwidth,clip}
\caption{
Evolution of the energy bands for selected numbers $(N_{y'},N_{z'})$ of oscillator states
in the matrix discretization  (columns) and aspect ratios $a$ (rows).
The gray colour results are qualitative, indicating that the corresponding energy regions 
are full of bands. The insets in the rightmost columns show the details of those  
dense band distributions. Parameters: $B=0$, ${\cal R}_{z'}=2.6\hbar\omega_0\ell_0$,
${\cal R}_{y'}=0$,
growth direction $(001)$
and wire orientation $(110)$.}
\label{figpband}
\end{figure}

\end{widetext}

Figure \ref{figpband} displays the evolution  
of the band structure when
a) increasing $N_{y'}$ and $N_{z'}$ sequentially from left to right panels; and
b) increasing the deformation degree $a$  from top 
to bottom panels.
The right column shows results that are very close to physical convergence.
Looking at the successive band crossings at $q=0$, 
we notice that the truncation to $(N_{y'},N_{z'})=(1,1)$ grossly overestimates 
the energy separation between pairs of bands in all cases. It is remarkable
that for increasing flatness degree the $(1,1)$ truncation deviates
more and more from the right column. This is a consequence of the 
intersubband couplings induced by the kp and Rashba Hamiltonians: at least a few
bands in the shallow oscillator ($y'$) are essential even for large $a$'s.

More reasonable results are found for $(N_{y'},N_{z'})=(10,1)$, although the 
differences with the $(10,10)$ basis are still large quantitatively. 
In this case, however, increasing $a$ improves the quality of the description
since intersubband coupling is allowed at least in $y'$ direction.
Finally, the $(10,2)$ results are close to the converged ones and only the insets reveal
that sizeable differences are present at intermediate or low values
of $a$. These differences are small in the behavior of the upper bands
and become more and more important as the energy is reduced.
From this analysis we conclude that for our present purpose, namely the description 
of magneto-g-factors of several successive conductance steps, it is essential to include
enough oscillator bands in both $y'$ and $z'$ oscillators.

\end{document}